\documentclass[preprint,pteplogo]{ptephy_v2}

\preprintnumber{XXXX-XXXX} 
\usepackage{hyperref}

\begin{document}

\title{Three-dimensional recoil-electron reconstruction using combined optical imaging and waveform readout for electron-tracking Compton cameras}


\author{Tomonori Ikeda}
\affil{National Institute of Advanced Industrial Science and Technology (AIST), National Metrology Institute of Japan (NMIJ),  1-1-1 Central 3, Umezono, Tsukuba, Ibaraki, 305-8563, Japan \email{tomonori.ikeda@aist.go.jp}}

\author{Tatsuya Sawano}
\affil{
 Graduate School of Natural Science and Technology, Kanazawa University, Kakuma, Kanazawa, 920-1192, Ishikawa, Japan
}

\author{Naomi Tsuji}
\affil{
Institute for Cosmic Ray Research, University of Tokyo, Kashiwanoha 5-1-5, Kashiwa, Japan
}

\author{Yoshitaka Mizumura}
\affil{
 Institute of Space and Astronautical Science, Japan Aerospace Exploration Agency, Yoshinodai 3-1-1, Chuou, Sagamihara, 252-5210, Kanagawa, Japan
}





\begin{abstract}
Accurate reconstruction of recoil-electron directions is critical for enhancing the point-spread function of electron-tracking Compton cameras (ETCCs) in gamma-ray imaging.
Although full three-dimensional (3D) readout systems achieve high-precision reconstruction, they are impractical for large-area detectors because of the enormous data volume.
This study proposes and demonstrates a practical alternative for inferring the 3D recoil-electron direction in Compton scattering.
This method combines a high-resolution two-dimensional optical image, a one-dimensional waveform signal, and a deep-learning-based method through simulations.
The proposed method achieved an angular resolution of approximately $44^\circ$ for the recoil-electron direction in the 40--50~keV range, corresponding to an improvement of a factor of about 1.3 compared with our previous strip-readout approach using pseudo-experimental data generated by Geant4 and MAGBOLTZ simulations for an argon-based gas time projection chamber. 
In addition,  the starting-point resolution of the electron track was improved over the previous method across the 5--50~keV electron energy range.
These results demonstrate that complementary information from the transverse image and longitudinal waveform can effectively recover the 3D track topology without requiring full 3D readout. 
The proposed approach provides a realistic pathway for improving ETCC imaging performance.
\end{abstract}


\subjectindex{xxxx, xxx}

\maketitle

\section{Introduction}
Observations of MeV and sub-MeV gamma rays enable a unique probe of astrophysical processes such as electron-positron annihilation in the Galactic center region~\cite{Siegert, Kierans_2020, Siegert_2020}, nucleosynthesis ~\cite{Diehl2013}, and non-thermal emission from compact objects~\cite{Chiaberge_2001}. 
In particular, the 511~keV annihilation line and the three-photon continuum from ortho-positronium decay provide important probes of the origin and propagation of Galactic positrons.
Despite their importance, observations in the MeV energy range remain limited because of the difficulties in achieving both high angular resolution and background suppression. 
Electron-tracking Compton cameras (ETCCs) have been developed to overcome these limitations by reconstructing the incident gamma-ray direction and resolving the full Compton-scattering kinematics, including information on the recoil-electron direction~\cite{TANIMORI2017}. 
In ETCCs, the point-spread function (PSF) strongly depends on the accuracy of the reconstructed recoil-electron direction~\cite{Tanimori2015}; thus, the improvement of the precision of recoil-electron reconstruction is essential.

Various approaches have been proposed for reconstructing electron tracks and recoil-electron directions in gaseous detectors ~\cite{Soffitta_2017, ODell_2019, Bellazzini_2003, DiMarco_2022, KITAGUCHI_2018, Li_2017, Adams_2019, Laura_2021, KITAGUCHI_2019, PEIRSON_2021}. 
In our previous study, we developed an ETCC equipped with a micro pixel chamber ($\mu$-PIC)~\cite{OCHI_2001,Takada2007} as the readout plane, and applied a deep learning (DL) method to reconstruct the recoil-electron direction from the measured two-dimensional (2D) track images~\cite{Ikeda_2021}.
The $\mu$-PIC is a strip readout detector with an intrinsic pitch of 400~$\mu$m, and the readout was performed by summing two channels, resulting in an effective pitch of 800~$\mu$m. 
Consequently, two projected images, $xz$ and $yz$, were obtained, where the longitudinal axis $z$ corresponds to the drift direction of the time projection chamber (TPC).
The $z$-axis information was reconstructed from the time-over-threshold (TOT) signal and timing information, which depends on the drift time of the electrons and the shaping time of the amplifiers.
Because the width of the TOT signal typically corresponded to several centimeters, the achievable resolution along the $z$-axis direction was limited, indicating that the accuracy of the $z$ determination, given a realistic drift velocity, was substantially poorer than that in the $xy$ plane.

Conversely, full three-dimensional (3D) voxel-based readout enables high-precision reconstruction when fine spatial information is available~\cite{Ghrear_2024}. 
However, implementing such a system in large-area detectors such as ETCCs is challenging because of the enormous data volume requirement.

These considerations motivate the development of a reduced-data approach for recovering 3D track information from reduced-dimensional measurements. 
In particular, a detector configuration combining one high-resolution 2D optical image with a one-dimensional (1D) waveform signal provides complementary information: 
the optical image provides precise transverse track morphology, whereas the waveform encodes the longitudinal structure through the drift time of ionization electrons. 
However,  no previous study has demonstrated a method for reconstructing the full 3D recoil-electron direction or the 3D track by integrating these two types of data modalities.

In this study, we propose and demonstrate a multistage DL framework that integrates a 2D optical image and 1D waveform data to reconstruct the 3D recoil-electron direction. 
Our aim is to improve the accuracy of the reconstructed 3D trajectory by combining complementary transverse and longitudinal information, rather than to improve the intrinsic spatial resolution of the detector. 

The remainder of this paper is divided into five sections. In Sec.~\ref{sec:data_preparation}, a pseudo-experimental dataset used to train the deep-learning model is introduced. 
In Sec.~\ref{sec:network_model}, three neural-network architectures are developed to extract representative track points, reconstruct the 3D track trajectory, and estimate the recoil-electron direction. 
In Sec.~\ref{sec:results}, the performance of the network model is evaluated, and the accuracy of the starting-point of the electron track and recoil-electron direction is compared with that of previous studies. 
The limitations of the proposed network model and future prospects are discussed in Sec.~\ref{sec:discussion}.
Finally, we summarize our findings in Sec.~\ref{sec:conclusion}.

\section{Data preparation} \label{sec:data_preparation}
\begin{figure}
    \centering
    \includegraphics[width=0.5\linewidth]{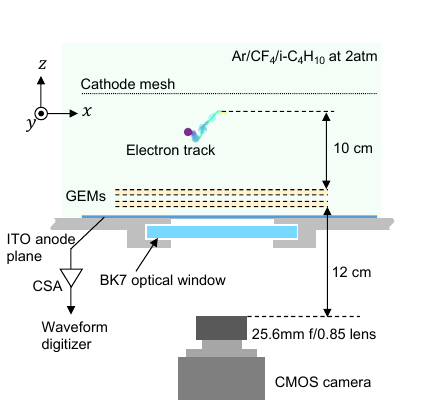}
    \caption{
Schematic view of the detector configuration assumed in the simulation. The TPC consists of stacked GEMs and a transparent ITO anode plane. A CMOS camera images the scintillation light emitted from the GEMs, while the charge signal is read out through the ITO anode.
}
\label{fig:geometry}
\end{figure}

In this section, we describe the pseudo-experimental dataset used to train the DL model.
Figure~\ref{fig:geometry} shows a schematic view of the detector configuration assumed in our simulation.
The TPC is filled with argon-based gas (95\%Ar + 3\%CF$_{4}$ + 2\%iso-C$_{4}$H$_{10}$) at a pressure of 2 atm, corresponding to the same configuration used in our previous work~\cite{Ikeda_2021} for comparison.
This gas mixture is known to provide a favorable balance between a high electron drift velocity and low electron diffusion~\cite{ABGRALL201125}.
The readout plane consists of a stack of gas electron multipliers (GEMs) and a transparent indium tin oxide (ITO) anode electrode~\cite{Brunbauer} deposited on a glass substrate. We define the $xy$ plane as parallel to the GEM surface and the $z$ axis as aligned with the electron-drift direction.

Ionization electrons produced by the recoil electron drift toward the GEMs under the applied electric field and are amplified through the GEM stack.
The avalanche process produces scintillation light with an emission spectrum peaked at approximately 620~nm~\cite{Kaboth_2008}.
The scintillation light is transmitted through the ITO anode, glass substrate and a BK7 optical window and is imaged onto a CMOS camera using a 25.6~mm f/0.85 lens.
The distance between the bottom GEM and the lens is 12~cm.
The electron-track image is projected onto the CMOS sensor with a magnification factor of 0.27.
In parallel with the optical imaging, the charge signal induced on the ITO anode is read out by a charge-sensitive amplifier (CSA) and recorded as a waveform.
The optical image therefore provides the two-dimensional track information in the $xy$ plane, while the charge waveform provides information along the electron-drift direction.
Electron tracking with optical-readout gas detectors and charge readout through a transparent ITO electrode has been demonstrated experimentally~\cite{Phan_2016, Araujo_2023, Fujiwara_2016}.
The baseline parameters of the detector configuration are summarized in Tab.~\ref{tab:params}.

\begin{table}[!t]
\caption{Baseline parameters used to generate
the pseudo-experimental data.}
\label{tab:params}
\centering
\begin{tabular}{|l|l|c|}
\hline
Component & Parameter & Value \\
\hline
Gas & Gas mixture &
Ar/CF$_4$/i-C$_4$H$_{10}$ (95/3/2) \\
 & Pressure & 2 atm \\
\hline
Electron transport & Drift velocity &
$38~{\rm mm/\mu s}$ \\
 & Transverse diffusion &
$0.26~{\rm mm/\sqrt{cm}}$ \\
 & Longitudinal diffusion &
$0.24~{\rm mm/\sqrt{cm}}$ \\
\hline
Ionization & $W$ value & $27.2~{\rm eV}$ \\
 & Fano factor & 0.20 \\
\hline
GEM & Gas gain, $G_{\rm gas}$ &
$2.0\times10^{4}$ \\
\hline
Optical readout 
 & Lens & EHD-25085-C \\
 & Working distance & $120~{\rm mm}$ \\
 & Focal length & $25.6~{\rm mm}$ \\
 & F-number & $0.85$ \\
 & Geometric acceptance, $f_{\rm geo}$ & $3.89\times10^{-3}$ \\
 & Optical magnification & $0.27$ \\ 
 & Photon yield, $\eta$ & 0.34 \\
 & Photon detection efficiency, $f_{\rm eff}$ &
$1.0\times10^{-3}$ \\
\hline
CMOS image & Pixel size &
$0.28\times0.28~{\rm mm^2}$ \\
 & Image size & $64\times64$ pixels \\
 & Readout noise, $\sigma_{\rm read}$ &
$0.43~e^{-}$ \\
\hline
Waveform 
& $z$-bin width & $0.19~{\rm mm}$ \\
& ENC of CSA, $\sigma_{\rm ENC}$ & $4200~e^{-}$ \\
& Gain of CSA & $1.0~{\rm V/pC}$ \\
& Decay time constant of CSA, $\tau_{\rm pre}$ & $140~\mu{\rm s}$ \\
\hline
\end{tabular}
\end{table}


\subsection{Monte Carlo simulation}\label{sec:monte}
\begin{figure}
\centering
\begin{minipage}{0.32\linewidth}
\centering
\raggedright
\textbf{(a)}\\
\includegraphics[width=\linewidth]{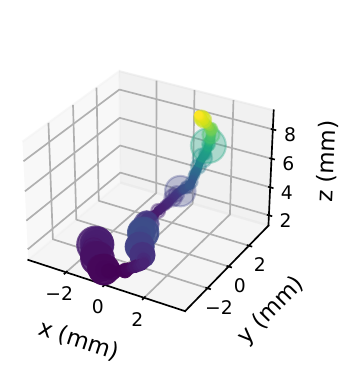}
\end{minipage}
\hfill
\begin{minipage}{0.32\linewidth}
\centering
\raggedright
\textbf{(b)}\\
\includegraphics[width=\linewidth]{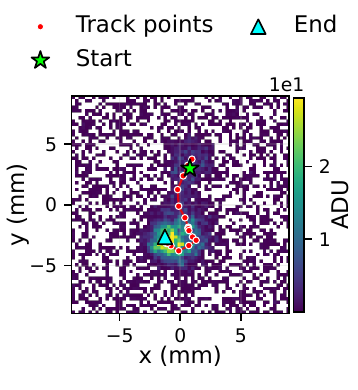}
\end{minipage}
\hfill
\begin{minipage}{0.32\linewidth}
\centering
\raggedright
\textbf{(c)}\\
\includegraphics[width=\linewidth]{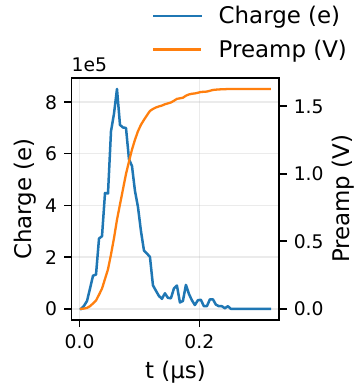}
\end{minipage}
\caption{
Example of pseudo-experimental data for a 49~keV electron.
(a) Three-dimensional electron trajectory obtained from the Geant4 simulation.
The marker size is proportional to the deposited energy along the track, and the color scale represents the $z$ coordinate.
(b) Simulated optical image projected onto the $xy$ plane after including diffusion and optical readout effects.
The color scale represents the detected signal in analog-to-digital units.
The red circles indicate the representative track points projected onto the $xy$ plane.
The green star and cyan triangle markers indicate the start and end points of the track, respectively.
(c) Simulated waveform at charge readout.
The blue curve represents the time profile of the collected charge, and the orange curve represents the corresponding preamplifier response.
}
\label{fig:event_sample}
\end{figure}

Electron tracks were simulated using Geant4 (version 11.0)~\cite{Geant4}.
Each primary electron was generated with a kinetic energy in the range of 5--50~keV, and the trajectory of the electron was tracked until its energy fell below the ionization threshold. 
Fig.~\ref{fig:event_sample} shows a typical event with an initial electron energy of 49~keV.
In this study, the primary electron is generated directly without simulating the Compton scattering process. Consequently, atomic de-excitation processes following the ionization of a bound electron, including the emission of Auger electrons and characteristic X rays, are not included in the simulation. A complete simulation of the ETCC detector response requires these secondary processes to be taken into account.

All events were generated at a fixed drift distance of 10~cm above the gas amplification region, with the initial directions uniformly distributed in the solid angle. A total of $2.5\times10^{6}$ events were uniformly produced in energy, and 20\% of the produced data were used as the validation data.

To incorporate the effect of charge transport to the gas amplification region, the diffusion of drift electrons was calculated using MAGBOLTZ~\cite{MAGBOLTZ}, which provides the longitudinal and transverse diffusion coefficients and the drift velocity.
The position of each ionized electron was smeared according to these diffusion parameters, thereby reproducing realistic charge spreading.
This hybrid Geant4–MAGBOLTZ simulation enables a consistent description of both microscopic energy loss and macroscopic charge transport in the TPC.
In our previous study, we demonstrated that a model trained on such simulated data can be directly applied to experimental data, confirming that the Geant4–MAGBOLTZ simulations accurately reproduce the performance of the ETCC~\cite{Ikeda_2021}.
In this study, we adopt a drift velocity of 3.8~cm/$\mu\rm{ s}$. The transverse and longitudinal diffusion coefficients are set to 0.26~$\rm{mm/\sqrt {cm}}$ and 0.24~$\rm{mm/\sqrt {cm}}$, respectively. 
These values correspond to those in a previous study~\cite{Ikeda_2021}.

\subsection{Optical image generation}
\label{sec:image_data}
Figure~\ref{fig:event_sample}~(b) shows an optical image of the scintillation light produced during gas amplification and projected onto the $xy$ plane.
For a single drift electron arriving at the amplification region, the expected number of detected photons, $N_{\rm det}$, is parameterized as follows:
\begin{equation}\label{eq:n_deg}
  N_{\rm det} = G_{\rm gas}\,\eta\,f_{\rm eff},
\end{equation}
where $G_{\rm gas}$ denotes the gas gain, $\eta$ represents the photon production ratio per avalanche electron, and $f_{\rm eff}$ denotes the overall photon detection efficiency on the optical readout.

The overall photon detection efficiency is parameterized as
\begin{equation}
    f_{\rm eff}
    = f_{\rm geo} T_{\rm ITO} T_{\rm win} T_{\rm lens}
      \epsilon_{\rm QE},
\end{equation}
where $f_{\rm geo}$ is the geometric acceptance of the lens,
$T_{\rm ITO}$, $T_{\rm win}$, and $T_{\rm lens}$ are the transmission
coefficients of the ITO anode, optical window, and lens, respectively,
and $\epsilon_{\rm QE}$ is the quantum efficiency of the CMOS sensor.
Considering the detector geometry, $f_{\rm geo}$ is estimated
to be $3.89\times10^{-3}$, assuming isotropic photon emission from the center of the GEM.
The transmission coefficients of the ITO anode and optical window are estimated to be approximately 0.83~\cite{Fujiwara_2016} and 0.90, respectively, with the latter corresponding to a typical commercial value. The resulting value of $f_{\rm geo}T_{\rm ITO}T_{\rm win}$ is approximately $2.9\times10^{-3}$.
The overall photon detection efficiency is expected to be below $2.9\times10^{-3}$ after accounting for additional losses due to the lens transmission and the quantum efficiency of the CMOS sensor. Because the wavelength-dependent transmission of the lens is unavailable, we conservatively adopt $f_{\rm eff}=1.0\times10^{-3}$ as the baseline value in the simulation. The adopted value is of the same order as that reported for an optical GEM readout system~\cite{Araujo_2023}.

We assume $G_{\rm gas}=2.0\times10^{4}$ and $\eta=0.34$~\cite{Kaboth_2008} for the baseline configuration.
The pixel size is set to $0.28~{\rm mm}\times0.28~{\rm mm}$, and each
event is represented by a $64\times64$ pixel image.
Gaussian readout noise is added independently to each pixel with
$\sigma_{\rm read}=0.43~e^{-}$, corresponding to the typical readout
noise of the Hamamatsu CMOS camera C15550-20UP.
No explicit photon-background subtraction or suppression is applied
to the images prior to the subsequent analysis.

In generating the optical images, the number of photons reaching the CMOS sensor was calculated using the simplified expression in Eq.~\ref{eq:n_deg}. Photon transport within the detector vessel, including reflections and scattering from the vessel walls and detector components, was therefore not modeled using ray tracing. Consequently, diffuse background photons arising from these processes are not included in the simulated images.


\subsection{Waveform data generation}
\label{sec:waveform_data}
The waveform was constructed from the simulated charge distribution along the drift $z$ direction. 
We assumed a non-pixelated charge readout, where the induced signal represents the total accumulated charge on the $xy$ plane as a function of time.
To emulate the electronic noise of the charge readout system, an equivalent noise charge (ENC) component was added to the collected charge. The ENC noise was modeled as Gaussian white noise with a root mean square value of $4200~e^{-}$.

The charge signal was then convolved with the response function of a charge-sensitive preamplifier, which was modeled using an exponential decay function with a decay constant of $\tau = 140~\mu{\rm s}$, and a charge-to-voltage conversion gain of $1~{\rm V/pC}$.
Finally, the waveform was digitized with a sampling time width of $5~{\rm ns}$ to reproduce the data acquisition system. 
The generated typical waveform is shown in Fig.~\ref{fig:event_sample}~(c).

\section{Network model} \label{sec:network_model}
In this section, we describe the neural network architecture used for the reconstruction of recoil-electron directions.
The proposed framework comprises three training stages: Stage~A, B, and C.
The Stage~A model estimates representative track points on the 2D optical image.
The Stage~B model reconstructs the 3D track trajectory using additional waveform information. Finally, the Stage~C model estimates the recoil-electron direction.
This multistage training strategy was implemented for two primary reasons.
First, some experimental configurations provide only 2D optical imaging data without waveform readout.
In such cases, the Stage~A model can still be applied to extract track features from the images.
Second, by introducing the intermediate reconstruction of the 3D track in Stage~B, the network is guided by a physically meaningful constraint before performing the final direction estimation.

\subsection{Stage A : Track point extraction}
\subsubsection{Architecture}
\begin{figure}
    \centering
    \includegraphics[width=1\linewidth]{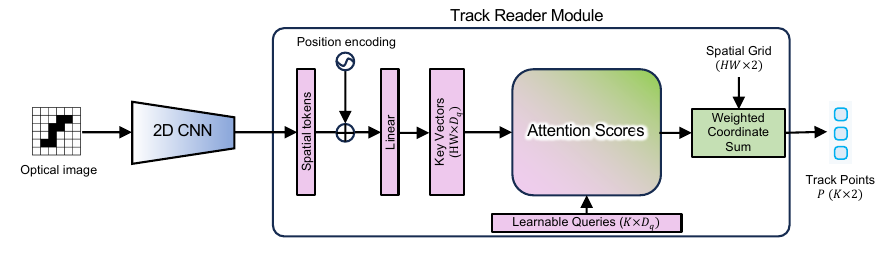}
    \caption{
    Architecture of Stage A. A 2D CNN extracts spatial features from the optical image, and a track reader module uses a query–attention mechanism to identify $K$ representative track points via a soft-argmax over the spatial grid.}
\label{fig:stage_a}
\end{figure}

Stage A extracts representative track points of the electron trajectory from a 2D optical image. The architecture of the Stage A is illustrated in Fig.~\ref{fig:stage_a}.

The input to the network is a single-channel optical image with a size of $1~\times~64~\times~64$. The image is first processed by a 2D convolutional neural network (CNN), which serves as a feature extractor.

The primary component of this stage is a track reader module, which extracts $K$ track-point coordinates $P_k=(x_k,y_k)$ from the CNN feature map using an attention-based mechanism.
This approach is conceptually inspired by the object-query mechanism used in the DETR framework~\cite{Carion_2020}, 
where learnable queries attend to spatial image features to identify objects. 
In our architecture, the learnable queries search for representative locations along the electron track. 
Attention weights are calculated from the CNN feature map, and the track point coordinates are obtained as a weighted spatial expectation.
Consequently, the network outputs $K$ track points, $P_k=(x_k,y_k)$ along the electron track in the $xy$ plane. 
Subsequently, these extracted track points are used in Stage B to reconstruct the 3D trajectory of the electron by incorporating waveform information. 

In this study, the number of track points was fixed at $K=32$.
The maximum track length is about 20~mm under the simulated conditions. Thus, $K=32$ representative points correspond to an average spacing of approximately 0.6~mm. In principle, a larger value of $K$ may be beneficial for longer tracks or for resolving finer structures. However, a larger $K$ increases the memory usage and computation required by the subsequent cross-attention layers of Stang~B. We used an NVIDIA GeForce RTX 3060 GPU with 12 GB of memory, and $K=32$ allowed stable training within the available GPU memory.

The trainable parameters include the convolutional kernels and batch-normalization affine parameters in the 2D CNN, the $K$ query embeddings, and the linear projections used to construct the query and key vectors. These parameters are optimized by backpropagation.



\subsubsection{Loss function}
To train Stage~A, the ground-truth track points were calculated by introducing a non-uniform sampling density along the track, enabling morphologically informative regions to be sampled more densely. 
The density function is defined as follows: 
\begin{equation}
g(s)=\epsilon_g+\lambda_q\,\tilde{q}(s)+\lambda_\kappa\,\tilde{\kappa}(s),
\end{equation}
where $s$ denotes the arc-length coordinate along the electron trajectory, $\tilde{q}(s)$ denotes the normalized local charge density and $\tilde{\kappa}(s)$ denotes the normalized curvature of the trajectory. 
The coefficients $\lambda_q$ and $\lambda_\kappa$ denote weighting parameters. Here, $\epsilon_g$ provides a small baseline density to ensure that the points are distributed along the entire trajectory.

To evaluate the normalized local charge density and the normalized curvature of the trajectory, the track was uniformly resampled into \(J=800\) points along the arc-length coordinate. The \(j\)-th resampled position is defined as,
\[
s_j=j\Delta s,\qquad \Delta s=\frac{L}{J-1},
\]
where $L$ is the total arc length of the trajectory projected onto the $xy$ plane, and $\Delta s$ is the spacing between adjacent resampled points. The relatively large value of $J$ was chosen to provide a sufficiently fine numerical discretization for evaluating the local energy deposition and curvature along the trajectory.
The local deposited energy at each resampled position was calculated within an arc-length window with a half-width of $h=2\Delta s$:
\begin{equation}
q(s_j)=\sum_{i:|s_i-s_j|<h}{E_i}.
\end{equation}
Here, \(i\) indexes the Geant4 energy-deposition hits, and \(E_i\) and \(s_i\) denote the deposited energy and the arc-length coordinate of the \(i\)-th hit, respectively.

Then, a logarithmic transformation is applied to compress its dynamic range:
\begin{equation}
q'(s_j)=\log(1+q(s_j)).
\end{equation}
Finally, $q'(s_j)$ is normalized independently for each event :
\begin{equation}
    \tilde{q}(s_j) = \frac{q_{\rm clip}(s_j)-q_5}{q_{95}-q_5},
\end{equation}
where $q_{95}$ and $q_5$ are 95th and 5th percentiles values of the $q'(s_j)$ distribution in each event.
The percentile clipping was introduced to reduce the influence of extreme energy-deposition values, which
would otherwise dominate the dynamic range.
The clipped charge quantity $q_{\rm clip}(s)$ is defined by,
\begin{equation}
    q_{\rm clip}(s_j) = \min [ \max (q'(s_j), q_5), q_{95}].
\end{equation}

The curvature of the trajectory is defined as:
\begin{equation}
    \kappa(s_j) = \frac{1}{\Delta s}\cos^{-1}\bigg(\frac{\boldsymbol{a}_{j}\cdot\boldsymbol{b}_{j}}{|\boldsymbol{a}_{j}||\boldsymbol{b}_{j}|}\bigg),
\end{equation}
where $\boldsymbol{a}_j=\boldsymbol{r}_j-\boldsymbol{r}_{j-1}$, $\boldsymbol{b}_j=\boldsymbol{r}_{j+1}-\boldsymbol{r}_j$ and $\boldsymbol{r}_j$ denotes the position vector of the $j$-th resampled point.
The curvature is normalized independently for each event in the same manner as the local charge quantity. The clipped curvature is defined as
\begin{equation}
\kappa_{\rm clip}(s_j)=\min\left[\max\left(\kappa(s_j),\kappa_5\right),\kappa_{95}\right],
\end{equation}
where $\kappa_5$ and $\kappa_{95}$ are the 5th and 95th percentile values, respectively, of the curvature distribution in each event. The normalized curvature is then given by
\begin{equation}
    \tilde{\kappa}(s_j)=\frac{\kappa_{\rm clip}(s_j)-\kappa_5}{\kappa_{95}-\kappa_5}.
\end{equation}

The $K$ ground-truth points were then determined by equal-spacing in the cumulative density as follows: 
\begin{equation}
G(s)=\int_0^{s} g(s')\,ds'.
\end{equation}
This effectively allocates more points to regions with higher charge density or larger curvature. 
The resulting coordinates define the ground-truth track points $P_k^{\mathrm{true}}$ used to train the network.

Because the ordering of the predicted track points was not predetermined, we employed a set-based loss function based on the Chamfer distance. 
The loss function is defined as follows:
\begin{equation}
L_{\mathrm{Chamfer}} =
\frac{1}{K}\sum_{k=1}^{K}
\min_{j}\|P_k^{\mathrm{pred}} - P^{\mathrm{true}}_j\|^2
+
\frac{1}{K}\sum_{j=1}^{K}
\min_{k}\|P^{\mathrm{true}}_j - P_k^{\mathrm{pred}}\|^2 ,
\end{equation}
where $P_k^{\mathrm{pred}}$ denotes the k-th predicted track point.

\subsection{Stage B : Waveform information incorporation for 3D track reconstruction}
\subsubsection{Architecture}
\begin{figure}
    \centering
    \includegraphics[width=1\linewidth]{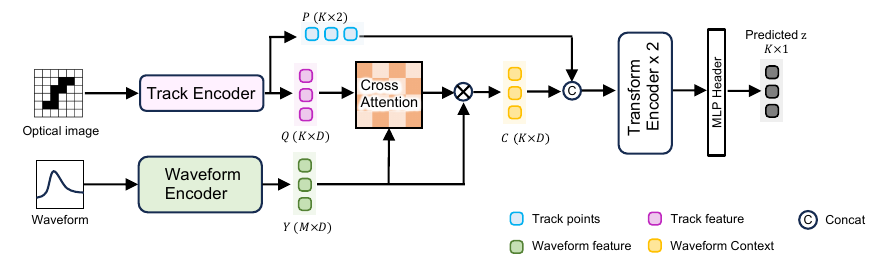}
    \caption{
    Architecture of Stage~B. 
    Track points extracted from the optical image act as queries, and the waveform signal is encoded into temporal tokens. 
    Cross-attention aligns the track points with the waveform tokens to produce context features that are processed using a transformer. An MLP header is then used to predict the longitudinal coordinate $z$.
    }
    \label{fig:stage_b}
\end{figure}

Stage~B extends the predicted 2D track points obtained in Stage~A by incorporating the waveform information to reconstruct the longitudinal coordinate $z_{k}$ of each track point.
The Stage~B architecture is illustrated in Fig.~\ref{fig:stage_b}. 
The architecture is based on a query-token alignment scheme: the $K$ track points extracted from the optical image act as queries, whereas the waveform is encoded as a sequence of tokens along the drift direction. 
Here, the term ``query-token'' follows the standard terminology used in Transformer-based machine learning architectures.

In the track encoder module, the query vector $Q_k$ associated with the spatial image feature is calculated.
The optical image is processed by a track encoder, which includes the track reader module introduced in Stage~A.
The track points $P_k$ are obtained by the track reader module.
In addition, the spatial features sampled at $P_k$ are embedded in tokens $X_k \in \mathbb{R}^{D}$, where $D$ represents the dimensionality of the embedding space.
The intensity feature at each track point is encoded as an additional auxiliary feature.
The resulting query vector used in the attention module is therefore constructed as follows:
\[
Q_k = \mathrm{Proj}([X_k, I_k, P_k]),
\]
where $I_k$ denotes the intensity feature of CMOS images from the track encoder.

In the waveform encoder, the waveform signal is encoded using a 1D convolutional encoder into a sequence of $M$ waveform tokens $Y_m \in \mathbb{R}^{D}$. Here, $Y_m$ does not correspond to an individual pulse or a Fourier component and does not have a direct physical unit.
A learnable positional encoding is added to the waveform tokens to provide the network with information about the temporal ordering of the waveform samples.

Cross-attention from each spatial token $Q_k$ to the waveform tokens $Y_m$ is used to identify waveform features associated with the $k$-th track point.
The corresponding attention weights are calculated as,
\begin{equation}
A_{k m} =
{\rm Softmax}
\left(
\frac{Q_k \cdot Y_m}{\sqrt{D}}
\right),
\end{equation}
where $A_{km}$ denotes the attention weight from the $k$-th spatial token to the $m$-th waveform token.

Using this alignment matrix, a waveform context vector is calculated as follows:
\begin{equation}
C_k = \sum_m ^{M} A_{km} Y_m .
\end{equation}
The context vectors are then concatenated with the track points and processed using a transformer operating along the sequence of $K$ points.
This stage allows the network to model correlations among neighboring track points.
Finally, the longitudinal coordinate $z$ is predicted using a multilayer perceptron (MLP) applied to the output features of the transformer.

The track encoder module based on the track reader module of Stage~A was initialized from the trained Stage A model and kept fixed throughout training. The remaining modules are trained in Stage B, including the track-token embedding networks, waveform encoder and its positional embeddings, cross-attention, Transform encoder, and $z$ prediction heads.


\subsubsection{Loss function}
The total loss function was defined as follows:
\begin{equation}
L_{\mathrm{StageB}}
=
L_{z}
+
\lambda_{\mathrm{smooth}} L_{\mathrm{smooth}}
+
\lambda_{\mathrm{spread}} L_{\mathrm{spread}},
\end{equation}
where $L_z$ denotes the point-wise regression loss for the longitudinal coordinate, $L_{\mathrm{smooth}}$ denotes a smoothness regularization term, and $L_{\mathrm{spread}}$ constrains the overall longitudinal extent of the reconstructed trajectory. 
The weighting coefficients \(\lambda_{\mathrm{smooth}}\) and \(\lambda_{\mathrm{spread}}\) were empirically set to \(1.0\times10^{-3}\) and \(0.4\), respectively, by examining the reconstruction performance during training.

The point-wise regression loss was defined using the smooth~L1 loss
(Huber loss) for the predicted $z_k$ and teacher longitudinal coordinates $z_k^{\mathrm{true}}$. This term directly penalizes the uncertainty in the reconstructed longitudinal coordinate of each track point.

To impose a smoothness constraint, a second-difference regularization was introduced.
The second finite difference was computed as follows:
\begin{equation}
\Delta^2 z_k
=
z_{k+2} - 2z_{k+1} + z_k.
\end{equation}
The corresponding smoothness loss was defined as follows:
\begin{equation}
L_{\mathrm{smooth}}
=
\frac{1}{K-2}
\sum_{k=1}^{K-2}
(\Delta^2 z_k)^2 .
\end{equation}
This term suppresses unphysical point-to-point fluctuations in the predicted longitudinal coordinates and promotes a smooth 3D trajectory.

The spread regularization was defined as follows:

\begin{equation}
L_{\mathrm{spread}}
=
\left[
\ln(s_{\mathrm{pred}}+\epsilon)
-
\ln(s_{\mathrm{true}}+\epsilon)
\right]^2 ,
\end{equation}

where $s_{\mathrm{pred}}$ and $s_{\mathrm{true}}$ denote the standard deviations of the predicted and ground-truth longitudinal coordinates, respectively; $\epsilon$ denotes a small constant introduced for numerical stability.

\subsection{Stage C: Direction estimation}\label{sec:Stage_C}
\begin{figure}
    \centering
    \includegraphics[width=1\linewidth]{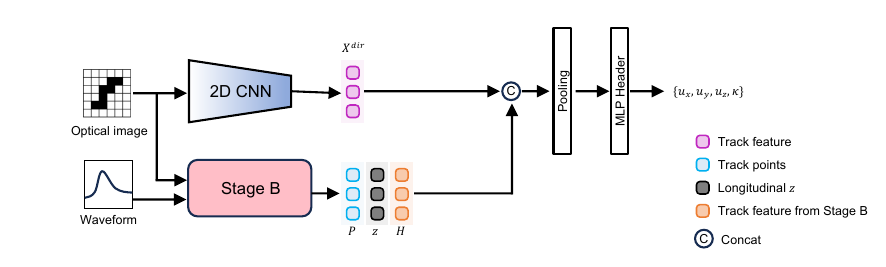}
    \caption{
    Architecture of the Stage~C. The frozen Stage~B backbone outputs the reconstructed track points $P_k$, longitudinal coordinates $z_k$, and latent track features $H_k$. These are concatenated with direction-sensitive image features and pooled to form an event-level representation, which is used to predict the recoil direction $\mathbf{u}$ and concentration parameter $\kappa$.
    }
    \label{fig:stage_c}
\end{figure}

Stage~C (Fig.~\ref{fig:stage_c}) performs the final recoil-electron direction estimation using the reconstructed 3D track obtained in Stages~A and B.

The initial electron direction is defined in the detector coordinate system introduced in Fig.~1. The unit direction vector is written as
\begin{equation}
\mathbf{u} =(\sin\theta\cos\phi,\, \sin\theta\sin\phi,\, \cos\theta),
\end{equation}
where $\theta$ is the polar angle and $\phi$ is the azimuthal angle in the $xy$ plane.

The networks trained in Stages~A and B are reused as a frozen backbone, whose parameters are kept fixed during the Stage~C training.
This backbone outputs the predicted track points $P_k=(x_k,y_k)$, their reconstructed longitudinal coordinates $z_k$, and the latent track features $H_k$ produced by the transformer in Stage~B.

An additional 2D CNN is introduced to extract direction-sensitive features from the optical image.
Feature vectors are sampled at the reconstructed track point positions and embedded in direction tokens, $X_k^{\mathrm{dir}}$.
The extracted features are integrated in a token-pooling module.
Finally, an MLP is used to predict the recoil direction.

In addition to the unit direction vector, the network can optionally predict a concentration parameter $\kappa$, which represents the uncertainty of the predicted direction.
This formulation follows the probabilistic directional regression framework proposed by Ghrear et al.~\cite{Ghrear_2024}, where the predicted direction is modeled using a von Mises–Fisher (vMF) distribution on the unit sphere $S^2$.
A larger $\kappa$ corresponds to a distribution more strongly concentrated around $\mathbf{u}$, indicating a higher confidence in the predicted direction, whereas a smaller $\kappa$ represents a broader directional distribution and hence a larger directional uncertainty. Therefore, $\kappa$ can be used to quantify the confidence in the reconstructed electron direction on an event-by-event basis.

During training, only the additional 2D CNN, direction-token embedding network, attention-pooling module, and the final prediction head for \(\mathbf{u}\) and \(\kappa\) were optimized.

\subsubsection{Loss function}
As mentioned in Sec.~\ref{sec:Stage_C}, we adopt the vMF distribution, which is a natural analogue of the Gaussian distribution on $S^2$ and has been used for directional prediction~\cite{Ghrear_2024}.
The probability density function is expressed as follows:
\begin{equation}
p(\mathbf{u}_i^{\mathrm{true}} \mid \mathbf{u}_i^{\mathrm{pred}}, \kappa_i)
=
\frac{\kappa_i}{4\pi \sinh \kappa_i}
\exp\!\left(\kappa_i \, \mathbf{u}_i^{\mathrm{pred}} \cdot \mathbf{u}_i^{\mathrm{true}}\right),
\end{equation}
where $\mathbf{u}_i^{\mathrm{pred}}$ and $\mathbf{u}_i^{\mathrm{true}}$ denote the predicted and ground-truth unit vector for the $i$-th event, respectively.
Maximizing this likelihood is equivalent to minimizing the negative log-likelihood (NLL),
\begin{equation}
\mathcal{L}_{\mathrm{dir}}
=
-\frac{1}{N}
\sum_{i=1}^{N}
\left[
\ln\!\left(\frac{\kappa_i}{4\pi \sinh \kappa_i}\right)
+
\kappa_i \, (\mathbf{u}_i^{\mathrm{pred}} \cdot \mathbf{u}_i^{\mathrm{true}})
\right].
\label{eq:stageC_vmf_nll}
\end{equation}

In practice, numerical approximations were introduced for the stable evaluation of the normalization term.
For small $\kappa$, the logarithmic normalization factor was approximated using a truncated Taylor expansion, whereas for larger $\kappa$ an asymptotic form was used, as described in Ref.~\cite{Ghrear_2024}.
This approach avoids the numerical instability associated with directly evaluating $\sinh \kappa$ and retains the probabilistic interpretation of the loss function.

In addition, we introduced a regularization term that suppresses artificial non-uniformities in the predicted angular distribution.
We penalized low-order Fourier modes in the angular space of the azimuthal angle $\phi$ and $\mu=\cos\theta$.
To construct the periodic variables of $\phi \in [0,2\pi)$ and $\psi \in [0,2\pi)$, we defined
\begin{equation}
\psi = \pi(\mu + 1).
\end{equation}
Here, \(\psi\) is not a physical direction angle but an auxiliary periodic variable introduced only for the uniformity regularization.
The loss function for uniformity is defined as follows:
\begin{equation}
\mathcal{L}_{\mathrm{uni}}
=
\sum_{(p,q)\neq(0,0)}
\left[
\left\langle \cos(p\phi + q\psi) \right\rangle^2
+
\left\langle \sin(p\phi + q\psi) \right\rangle^2
\right],
\label{eq:joint_fourier_loss}
\end{equation}
where $\langle \cdots \rangle$ indicates the average over a mini-batch. Furthermore, $p$ and $q$ denote the Fourier coefficients with respect to $\phi$ and $\psi$, respectively.
The vanishing of all Fourier coefficients except for the zeroth mode implies a uniform distribution. Therefore, minimizing Eq.~(\ref{eq:joint_fourier_loss}) suppresses systematic biases in the direction distribution.

The total loss used for training is then given by
\begin{equation}
\mathcal{L}_{\mathrm{total}}
=
\mathcal{L}_{\mathrm{dir}}
+
\lambda_{\mathrm{uni}} \, \mathcal{L}_{\mathrm{uni}},
\label{eq:total_loss_stageC}
\end{equation}
where $\lambda_{\mathrm{uni}}$ controls the strength of uniformity regularization.
Practically, this term was introduced progressively during training.
In the early stage of optimization, the network first learns the primary directional regression task using $\mathcal{L}_{\mathrm{dir}}$. After the coarse directional structure has been learned, the uniformity regularization term is gradually increased in a stepwise manner.

\section{Results} \label{sec:results}
In this section, the performance of the proposed multistage DL framework is presented. The performance of each stage is quantitatively evaluated and representative event displays are provided to illustrate the reconstruction behavior of the model.
\subsection{Result of Stage A}
\subsubsection{Evaluation metric}\label{sec:eval_stageA}
The \textit{covered arc fraction} was used to evaluate the extent to which the predicted track points cover the electron trajectory.
This metric evaluates the fraction of the true track arc length that lies within a predefined spatial distance threshold from at least one predicted track point.

The covered arc fraction is defined as follows:
\begin{equation}
f_{\mathrm{arc}} =
\frac{\int I(d(s) < d_{\mathrm{thr}}) ds}{\int ds},
\end{equation}
where $I$ represents the indicator function, $d_{\mathrm{thr}}$ denotes a spatial threshold, $s$ represents the arc-length coordinate along the ground-truth trajectory, and $d(s)$ denotes the distance between the true trajectory and the nearest predicted point.
A $f_{\mathrm{arc}}$ value of 1 indicates that the predicted points cover the entire track trajectory within the spatial threshold.

\begin{figure}
\centering
\begin{minipage}{0.49\linewidth}
\centering
\raggedright
\textbf{(a)}\\
\includegraphics[width=\linewidth]{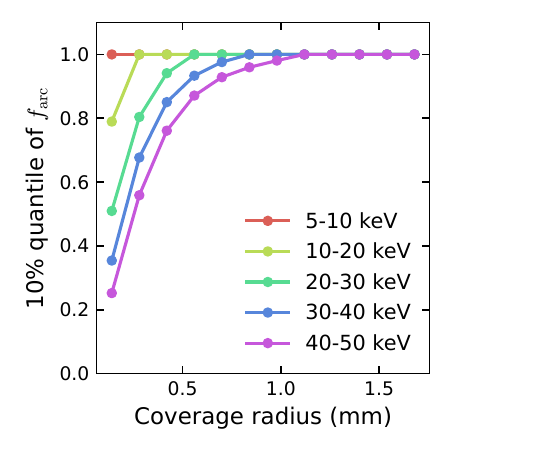}
\end{minipage}
\hfill
\begin{minipage}{0.49\linewidth}
\centering
\raggedright
\textbf{(b)}\\
\includegraphics[width=\linewidth]{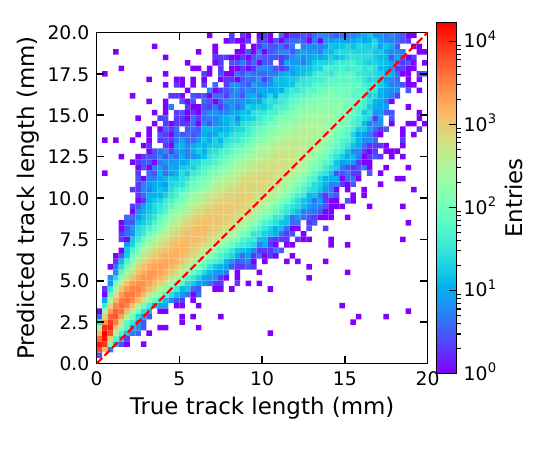}
\end{minipage}

\caption{
Performance metrics of Stage~A.
(a) Relation between the spatial threshold $d_{\mathrm{thr}}$ (coverage radius) and the 10\% quantile of the covered arc fraction $f_{\mathrm{arc}}$ for different energy ranges.
(b) Correlation between the true and predicted track length estimated from the extracted track points. The red dashed line represents the ideal relation $y=x$.
}
\label{fig:metrics_2D}
\end{figure}

For each $d_{\mathrm{thr}}$ value, the distribution of $f_{\mathrm{arc}}$ was calculated over the validation dataset, and the 10\% quantile of the distribution was extracted.
Figure~\ref{fig:metrics_2D}~(a) shows the obtained relation between the spatial threshold (coverage radius) $d_{\mathrm{thr}}$ and the 10\% quantile of $f_{\mathrm{arc}}$.
The 10\% quantile across all energy ranges reached nearly complete coverage at a $d_{\mathrm{thr}}$ threshold of $\sim 1~\mathrm{mm}$.
This indicates that the extracted track points follow the geometric structure of the electron trajectory with a spread of approximately 1~mm.

Figure~\ref{fig:metrics_2D}~(b) shows the correlation between the true 2D track length and the predicted 2D track length.
The obtained distribution is broadly aligned along the diagonal line corresponding to $y=x$, indicating that the extracted points preserve the overall geometry of the electron trajectory, particularly for longer tracks at higher energies.
In contrast, a systematic discrepancy was observed in the short track-length region, where the predicted track length tends to be slightly overestimated.
This tendency reflects the intrinsic difficulty in reconstructing short electron tracks.
Overall, these results indicate that the Stage~A network successfully extracts the global structure of the electron track, while exhibiting a modest bias in the reconstruction of short tracks.

\subsubsection{Event display}\label{sec:event_display_stageA}
\begin{figure}
\centering
\includegraphics[width=1\linewidth]{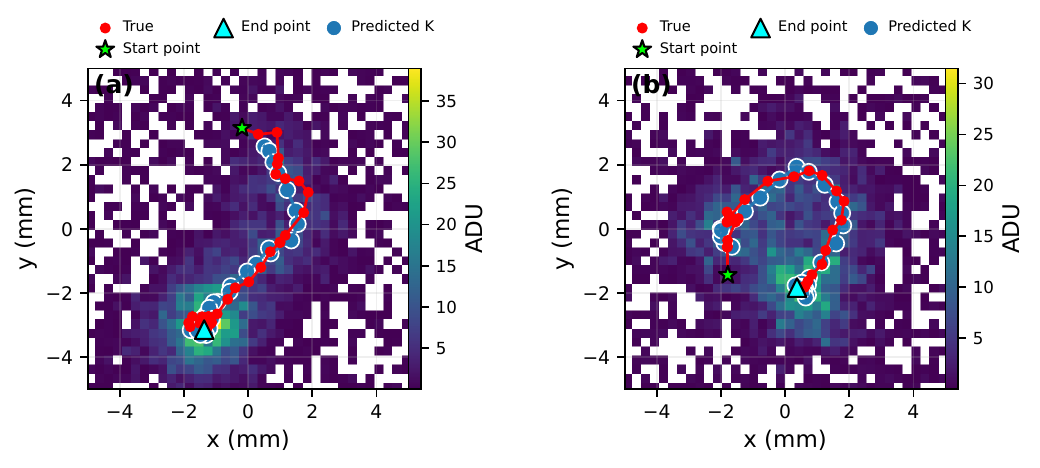}
\caption{
Examples of reconstructed track points extracted by the Stage~A network for electrons in the 40--50~keV energy range.
The background color map shows the optical image of the electron track.
The red curve represents the ground-truth trajectory projected onto the $xy$ plane, and the blue markers indicate the representative track points predicted by the network.
}
\label{fig:stageA_event}
\end{figure}

Typical event displays are shown in Fig.~\ref{fig:stageA_event}.
The predicted track points follow the electron trajectory.
These results demonstrate that the network successfully identifies representative locations along the projected trajectory using the optical image information.

\subsection{Results of Stage B}
\subsubsection{Evaluation metrics}
\begin{figure}
\centering
\begin{minipage}{0.49\linewidth}
\centering
\raggedright
\textbf{(a)}\\
\includegraphics[width=\linewidth]{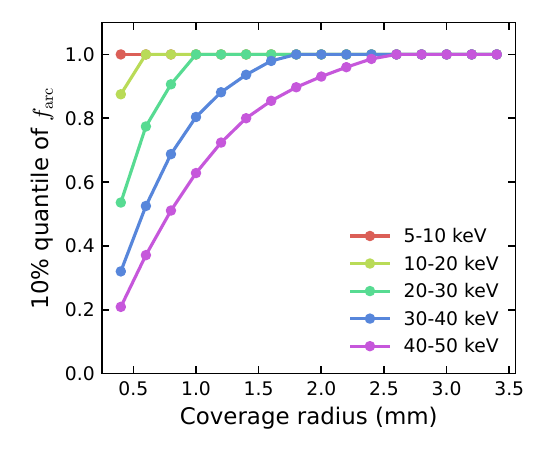}
\end{minipage}
\hfill
\begin{minipage}{0.49\linewidth}
\centering
\raggedright
\textbf{(b)}\\
\includegraphics[width=\linewidth]{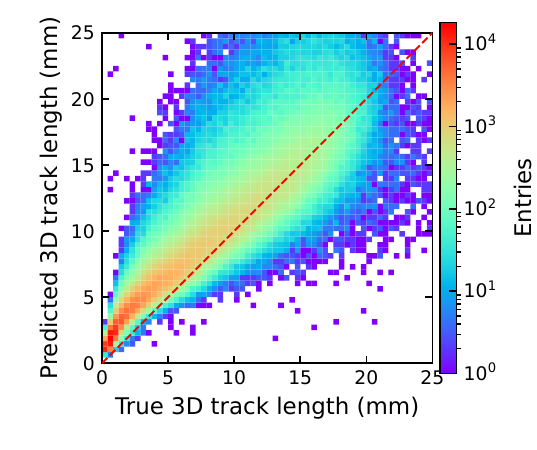}
\end{minipage}
\caption{
Performance of the Stage~B.
(a) Relation between the spatial threshold $d_{\rm thr}$ (coverage radius) and the 10\% quantile of the covered arc fraction $f_{\rm arc}$ for different energy ranges.
(b) Correlation between the true 3D track length and the predicted track length reconstructed from the track points.
The red dashed line represents the ideal relation $y=x$.
}
\label{fig:metrics_3D}
\end{figure}

The performance of Stage~B was evaluated using the covered arc fraction, as introduced in Stage~A.
The 10\% quantile of $f_{\rm arc}$ for each energy range was calculated using the same procedure as described in Sec.~\ref{sec:eval_stageA}.
Figure~\ref{fig:metrics_3D}~(a) shows the obtained relation between the spatial threshold and the 10\% quantile of $f_{\rm arc}$.
The 10\% quantile across all energy ranges reached nearly complete coverage at a $d_{\mathrm{thr}}$ threshold of $\sim 2~\mathrm{mm}$.

Figure~\ref{fig:metrics_3D}~(b) shows the correlation between the true 3D track length and the track length estimated from the reconstructed track points.
The distribution is broadly aligned along the diagonal ($y=x$), indicating that the reconstructed points preserve the global geometry of the electron trajectory in three dimensions.
A slight overestimation is observed in the short track-length region.
The distribution exhibits a larger spread than the 2D result, particularly at longer track lengths.
This indicates the additional uncertainty introduced in reconstructing the longitudinal $z$ component from the waveform signal.

\begin{figure}
    \centering
    \includegraphics[width=0.5\linewidth]{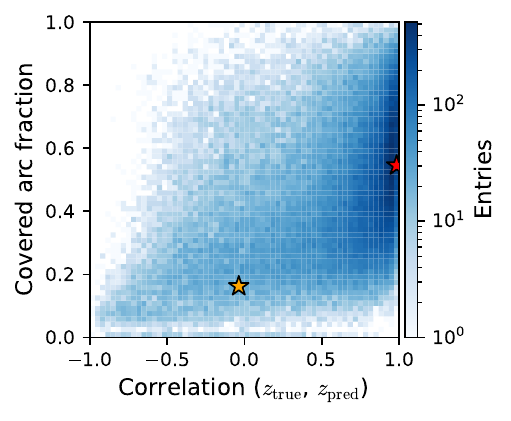}
    \caption{
Distribution of the correlation
between the true and predicted longitudinal coordinates
and the covered arc fraction for electrons in the 40--50~keV energy range
The red and orange stars indicate representative high-performance and low-performance events, respectively.
}
\label{fig:zcorr_vs_arc}
\end{figure}

To further investigate the performance of 3D track reconstruction, the correlation between the true and predicted $z$ coordinates for each event was calculated.
Figure~\ref{fig:zcorr_vs_arc} shows the 2D distribution of the covered arc fraction as a function of the correlation coefficient of $z$ coordinates for electrons in the 40--50~keV energy range.

A clear tendency was observed in which events with higher $z$-correlations achieve larger covered arc fractions.
In particular, the region near $\mathrm{correlation}(z_{\rm true}, z_{\rm pred}) \approx 1$ is dominated by events with high coverage, supporting the interpretation that accurate reconstruction of the longitudinal coordinate leads to a faithful recovery of the 3D electron trajectory.
Conversely, events with low correlation coefficients are concentrated in the region of a small covered arc fraction, indicating that failures in the reconstruction of the $z$ coordinate are a major factor limiting the overall track reconstruction performance.

\subsubsection{Event display}
\begin{figure}
\centering
\begin{minipage}{0.49\linewidth}
\centering
\raggedright
\textbf{(a)}\\
\includegraphics[width=\linewidth]{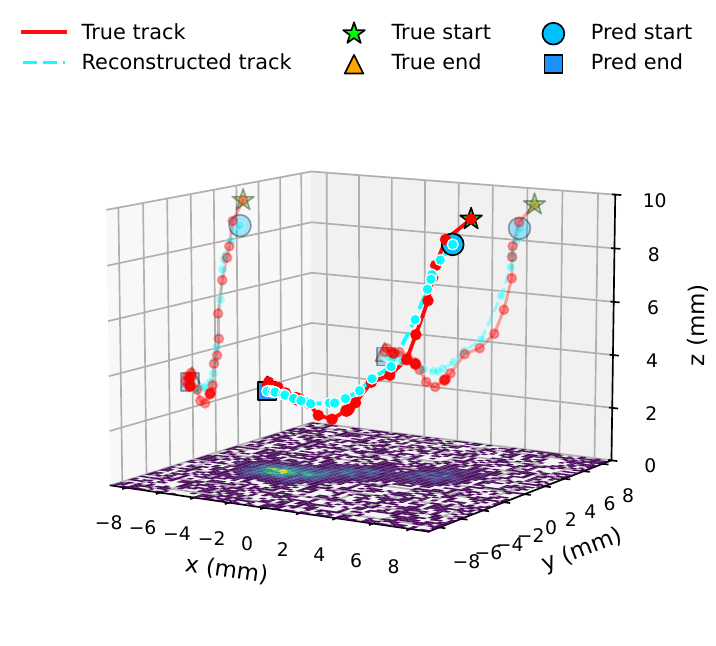}
\end{minipage}
\hfill
\begin{minipage}{0.49\linewidth}
\centering
\raggedright
\textbf{(b)}\\
\includegraphics[width=\linewidth]{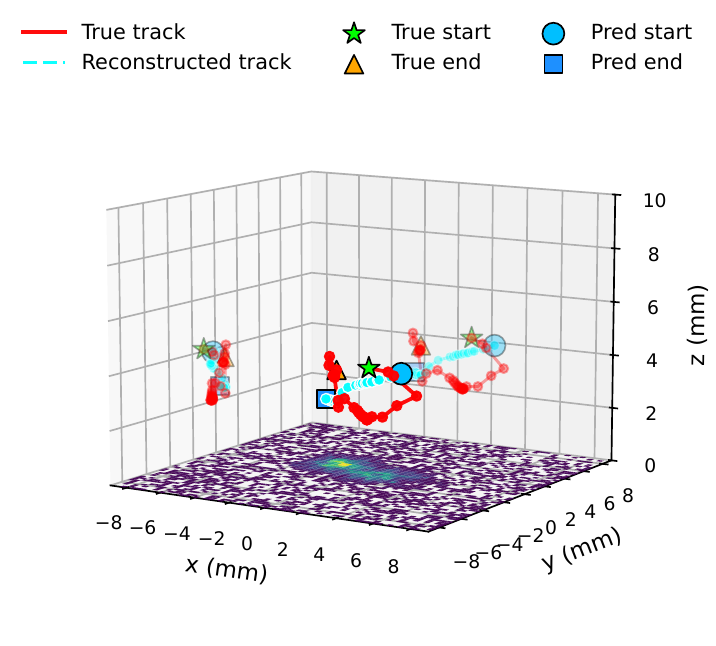}
\end{minipage}
\caption{
Event displays of 3D track reconstruction.
The red markers and solid line represent the true electron track obtained from the simulation, and the cyan dashed line shows the reconstructed trajectory predicted by the network.
The green stars and orange triangles denote the true start and end points of the track, respectively. The blue circle and square indicate the predicted start and end points, respectively. The color map on the bottom plane corresponds to the CMOS image of the track projected onto the $xy$ plane.
(a) and (b) Examples of high- and low-performance events, respectively.
}
\label{fig:event_display}
\end{figure}

To illustrate reconstruction behaviors, typical event displays are shown in Fig.~\ref{fig:event_display}.
Figure~\ref{fig:event_display}~(a) shows an example of reconstructed 3D trajectories selected from the high-performance event marked by the red star in Fig.~\ref{fig:zcorr_vs_arc}.
The reconstructed trajectory closely follows the true electron track, indicating that the network successfully recovers the 3D structure of the electron trajectory.

For comparison, a typical event selected from the low-performance event marked by the orange star in Fig.~\ref{fig:zcorr_vs_arc} is shown in Fig.~\ref{fig:event_display}~(b).
In this example, the electron trajectory exhibits a sharp bending along the $z$ direction.
Consequently, two distinct segments of the 3D track overlap in the projected $xy$ image.
Because these segments cannot be clearly separated in the CMOS image, the spatial correspondence between the projected track and the temporal timing structure of the waveform becomes ambiguous.
Under such conditions, the network fails to reconstruct the 3D track due to insufficient information.

\subsubsection{Scattering point resolution}
\begin{figure}
    \centering
    \includegraphics[width=0.5\linewidth]{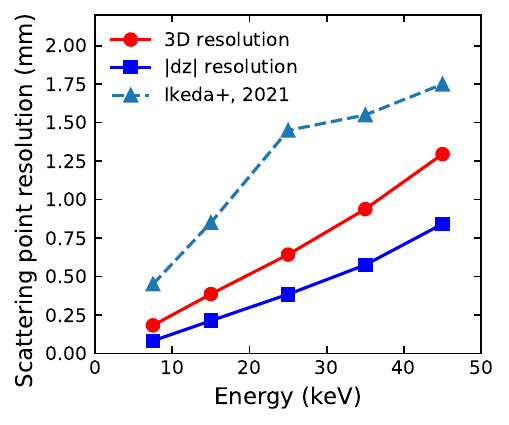}
    \caption{
Scattering point resolution as a function of electron energy.
The red circles show the 3D spatial resolution obtained in this
study, and the blue squares represent the resolution of the longitudinal
component $|dz|$.
The cyan dashed line represents the result reported by Ikeda et al.~(2021)
obtained using a strip-type gas detector with a machine learning-based
reconstruction method.}
    \label{fig:reso_scat}
\end{figure}

The scattering point resolution (starting-point resolution of the electron track) is defined as the median of the distribution of the spatial distance between the true and predicted scattering points.
Figure~\ref{fig:reso_scat} shows the scattering point resolution as a function of the electron energy.
The red and blue markers represent the 3D and the longitudinal component $|dz|$ spatial resolution of the scattering point obtained in this study.
For comparison, the cyan marker represents the scattering point resolution reported in the previous study by Ikeda et al.~\cite{Ikeda_2021},
who used a 2D strip-type gas detector $\mu$-PIC combined with a machine learning-based reconstruction method.

The proposed method achieves a better scattering point resolution over the entire energy range than the previous study.
This result indicates that the combination of the CMOS optical image and
the waveform signal provides sufficient information to reconstruct the
scattering point with high accuracy, even without using a strip-type charge readout detector.

\subsection{Results of Stage C}

\subsubsection{Correlation matrices}
\begin{figure}
    \centering
    \includegraphics[width=1\linewidth]{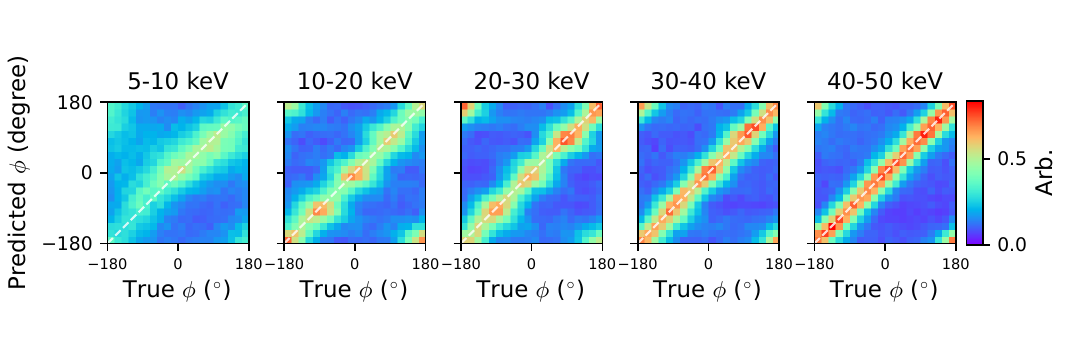}
    \caption{
    Correlation matrix between the true and predicted azimuthal angles $\phi$ for different electron energy ranges.
    }
    \label{fig:phi_correlation}
\end{figure}

Figure~\ref{fig:phi_correlation} shows the correlation matrix between the true and predicted azimuthal angles $\phi$ for the investigated electron energy ranges.
A clear diagonal structure is observed in all energies, indicating that the Stage C network successfully reconstructs the azimuthal component of the direction.
The spread reduction around the diagonal with increasing energy reflects the improved track reconstruction due to reduced multiple scattering and increased track length.

\begin{figure}
    \centering
    \includegraphics[width=1\linewidth]{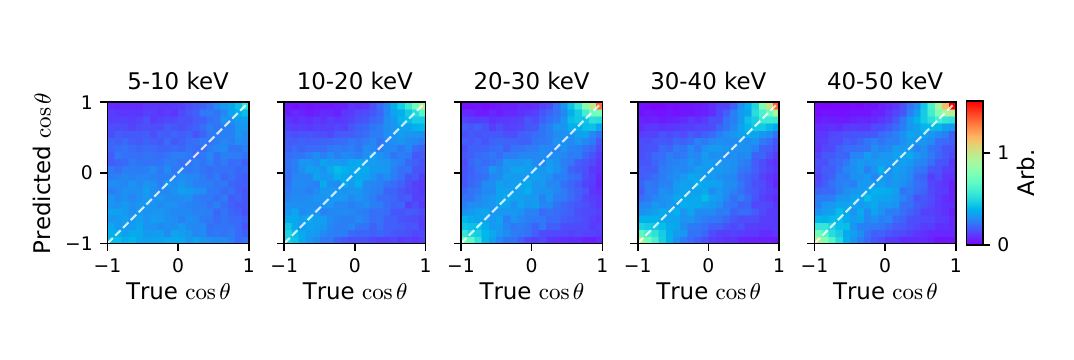}
    \caption{
    Correlation matrix between the true and predicted $\cos\theta$ for different electron energy ranges.
    }
    \label{fig:theta_correlation}
\end{figure}

Figure~\ref{fig:theta_correlation} shows the correlation matrix between the true and predicted cosine of the polar angle, $\cos\theta$.
In contrast to the $\phi$ correlation, the distribution in $\cos\theta$ is significantly broader, indicating that the reconstruction accuracy in the polar direction is degraded.
This behavior reflects the limited sensitivity of the waveform information to the fine $z$-structure of the track near the starting point.
However, a modest improvement is observed at higher energies, where longer track lengths provide additional constraints on the 3D geometry of the track.


\subsubsection{Angular resolution}
\begin{figure}
\centering
\begin{minipage}{0.49\linewidth}
\centering
\raggedright
\textbf{(a)}\\
\includegraphics[width=\linewidth]{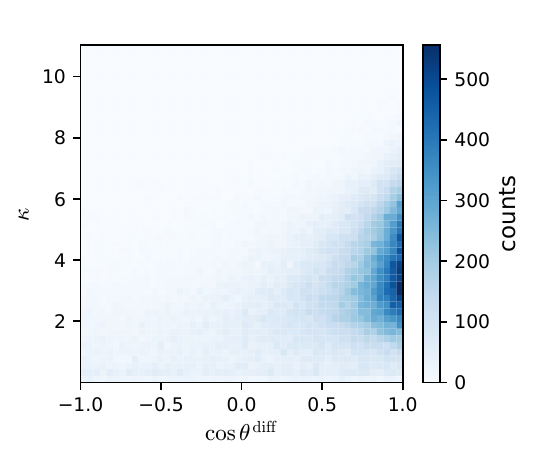}
\end{minipage}
\hfill
\begin{minipage}{0.49\linewidth}
\centering
\raggedright
\textbf{(b)}\\
\includegraphics[width=\linewidth]{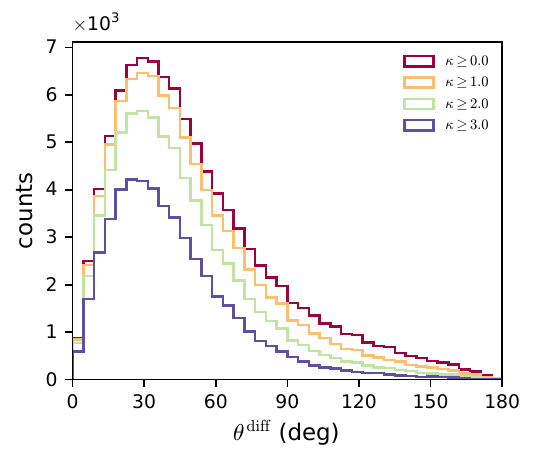}
\end{minipage}
\caption{
(a) Two-dimensional distribution of the concentration parameter $\kappa$ versus $\cos\theta^{\mathrm{diff}}$, where $\theta^{\mathrm{diff}}$ denotes the angular difference between the true and predicted directions in the energy range of 40--50~keV.
(b) Distributions of $\theta^{\mathrm{diff}}$ for different $\kappa$ thresholds in the energy range of 40--50~keV.
}
\label{fig:cos_dist}
\end{figure}

The angular difference is defined as follows:
\begin{equation}
\theta_i^{\mathrm{diff}} = \arccos(\mathbf{u}_i^{\mathrm{pred}} \cdot \mathbf{u}_i^{\mathrm{true}}).
\end{equation}
Figure~\ref{fig:cos_dist}~(a) shows the distribution of the angular difference and the concentration parameter $\kappa$ in the energy range of 40--50~keV.
Based on this distribution, the angular resolution was evaluated by applying thresholds on $\kappa$.
Figure~\ref{fig:cos_dist}~(b) shows the $\theta^{\rm diff}$ distributions for different $\kappa$ thresholds in the energy range of 40--50~keV.
As the $\kappa$ threshold increases, the fraction of mis-reconstructed events with opposite directions is significantly reduced.
This indicates that $\kappa$ effectively suppresses poorly reconstructed events and improves the angular resolution.

\begin{figure}
\centering
\begin{minipage}{0.49\linewidth}
\centering
\raggedright
\textbf{(a)}\\
\includegraphics[width=\linewidth]{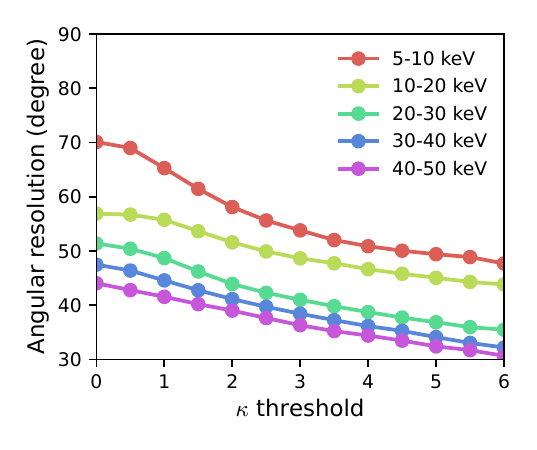}
\end{minipage}
\hfill
\begin{minipage}{0.49\linewidth}
\centering
\raggedright
\textbf{(b)}\\
\includegraphics[width=\linewidth]{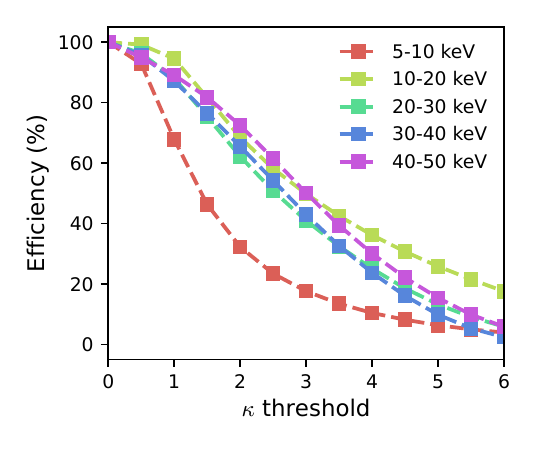}
\end{minipage}
\caption{
(a) Angular resolution as a function of the $\kappa$ threshold for different electron energy ranges.
(b) Efficiency as a function of the $\kappa$ threshold.
}
\label{fig:ang_reso_kappa}
\end{figure}

The angular resolution for each energy with different $\kappa$ thresholds is shown in Fig.~\ref{fig:ang_reso_kappa}~(a), while the corresponding selection efficiency is presented in Fig.~\ref{fig:ang_reso_kappa}~(b).
The angular resolution is defined as the median of the $\theta^{\mathrm{diff}}$ distribution, following the definition adopted by Ikeda et al.~\cite{Ikeda_2021}.
The directional dependence of the selection efficiency after the $\kappa$ selection is shown in Appendix~\ref{app:kappa_direction}.

\begin{figure}
    \centering
    \includegraphics[width=0.5\linewidth]{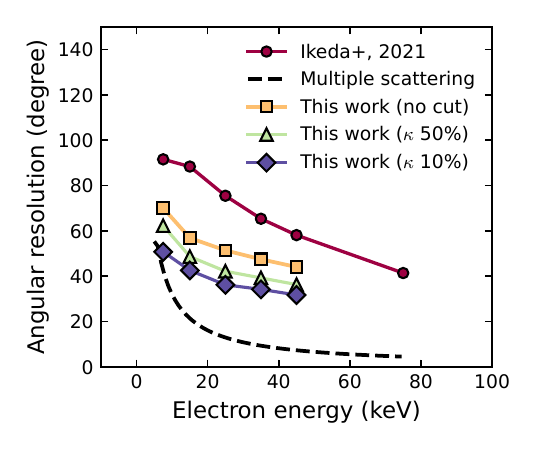}
    \caption{
    Angular resolution as a function of electron energy.
    The results without $\kappa$ selection (orange squares) and with $\kappa$ selections (green triangles for top 50\% and blue diamonds for top 10\%) are shown.
    The previous work by Ikeda et al.~(2021) (red circles) and the theoretical limit derived from multiple scattering for a 0.28~mm track length (black dashed line) are also described.
    }
    \label{fig:result_ang_reso}
\end{figure}

Finally, Fig.~\ref{fig:result_ang_reso} summarizes the angular resolution as a function of the electron energy.
The resolution improves monotonically with increasing energy.
Furthermore, the application of a $\kappa$ selection further enhances the angular resolution at the cost of reduced event statistics.
In the 40--50~keV range, the angular resolution reaches approximately $44^\circ$ without any $\kappa$ selection, outperforming the previous study~\cite{Ikeda_2021} by a factor of 1.3.
With a 50\% $\kappa$ selection, the angular resolution is further enhanced to approximately $36^\circ$, and with a 10\% $\kappa$ selection, it reaches approximately $32^\circ$.
These results demonstrate that the proposed reconstruction method provides intrinsically higher directional accuracy.

\section{Discussion} \label{sec:discussion}
We demonstrate that the 3D recoil-electron direction can be inferred from a hybrid detector comprising a high-resolution 2D optical image and a 1D waveform signal.
The success of the proposed reconstruction method can be explained by the complementary nature of the observables and their physical constraints.
The 2D optical image provides high-resolution information about the transverse morphology of the electron track, which contains physical properties such as track continuity, energy loss distribution, and multiple scattering. 
Although the optical image alone does not provide sufficient information to reconstruct the longitudinal (\(z\)) structure, these physical constraints limit the possible 3D configurations.
In contrast, the waveform signal provides temporal timing information, thereby encoding the longitudinal structure of the track. 

The cross-attention mechanism introduced in Stage~B plays a critical role in integrating these complementary observables. 
By learning the correlation between the spatial features in the optical image and the temporal timing features in the waveform, the Stage B network effectively performs a soft alignment between transverse and longitudinal information. 
This process resolves ambiguities in each modality and enables a 3D reconstruction of the electron track.
Consequently, the combination of physically constrained 2D morphology and temporally encoded longitudinal information allows the model to infer the 3D track structure with higher accuracy.

However, because the waveform contains only integrated timing information, the fine $z$-structure of the initial segment of the track cannot always be uniquely recovered.
This limitation is particularly important for events in which different parts of the 3D trajectory overlap in the projected $xy$ image, resulting in ambiguity in the correspondence between image and waveform features.
To resolve this ambiguity, additional information is needed.
We note that optical TPC concepts combining a CMOS camera with segmented strip readout have already been proposed~\cite{Araujo_2023}.
Such additional strip waveform information would help reduce the ambiguity in the longitudinal coordinate and is therefore expected to improve the reconstruction accuracy.

In the present simulation, the drift distance was fixed for all events.
In a realistic detector, the drift length varies event by event and affects electron diffusion, light yield, and the waveform shape.
Therefore, a more realistic treatment requires the incorporation of this variation into the simulation and training procedure. This issue will be addressed in future work.

In addition, the energy resolution of the gas detector is an important factor in Compton-event reconstruction, although the present model does not estimate the electron energy. Considering the energy resolution of the scintillator detector used as the absorber in an ETCC, the scintillator detector is expected to provide the dominant contribution to the uncertainty in the total-energy measurement. However, the energy resolution of the gas detector is not negligible. A quantitative evaluation of its impact on Compton-event reconstruction will be addressed in future work.

The present study focuses on single recoil-electron events. Longitudinal signals from multiple tracks can overlap in the waveform if their \(z\)-distributions overlap, making independent 3D reconstruction difficult because the charge signal is read out from a single non-pixelated anode electrode.

Finally, the results have important implications for designing ETCC detectors. 
The proposed method achieves better angular and scattering point resolutions than those obtained with conventional strip-based readout systems. 
The combination of a CMOS optical readout and waveform information provides a promising detector design alternative.
This approach is expected to endow ETCCs with enhanced imaging capability.
Especially, the concentration parameter $\kappa$ introduced in this study provides a physically meaningful event-by-event measure of reconstruction quality.
Although applying a $\kappa$ selection inevitably reduces the event statistics, it significantly improves the angular resolution by suppressing misreconstructed events.
As a result, the point spread function of an ETCC for gamma-ray imaging is improved.
Such a controllable trade-off parameter is highly advantageous for ETCC analyses, where event selection criteria can be tuned to optimize imaging performance or sensitivity depending on the scientific target.

\section{Conclusion}\label{sec:conclusion}

In this study, we demonstrated the feasibility of reconstructing the 3D recoil-electron direction using a hybrid detector that combines a high-resolution 2D optical image with a 1D waveform signal via simulations. 
A multistage DL framework was developed to integrate these complementary modalities and reconstruct the recoil-electron direction.
The proposed method achieved an angular resolution of approximately $44^\circ$ for the recoil-electron direction in the 40--50~keV energy range, corresponding to an improvement by a factor of 1.3 compared with our previous strip-based approach. 
In addition, the starting-point resolution of the electron track was improved across the investigated energy range. 
Furthermore, we demonstrated that event selection based on the concentration parameter $\kappa$ improves the angular resolution.
The proposed approach provides a practical and scalable alternative to conventional strip-based systems and offers a realistic pathway for improving the imaging capability of ETCCs.

\section*{Acknowledgment}
This study was supported by the Japan Society for the Promotion of Science (JSPS) KAKENHI Grants-in-Aid for Scientific Research (Grant Numbers 22J00064, 22KJ1766, and 24K00643), and a grant from the Precision Measurement Technology Promotion Foundation (PMTPF).


%

\vspace{0.2cm}
\noindent


\let\doi\relax


\bibliographystyle{ptephy}
\bibliography{main}

\appendix
\section{Directional dependence of the efficiency after $\kappa$ selection}
\label{app:kappa_direction}
We evaluated the relative selection
efficiency as a function of the true electron direction to examine whether the event selection based on $\kappa$ introduces a
direction-dependent acceptance.
For each bin of the true azimuthal angle $\phi$ or true $\cos\theta$,
the relative efficiency was calculated as the fraction of events
satisfying $\kappa \geq \kappa_{\rm th}$, normalized to the number of
events satisfying $\kappa \geq 0$ in the same directional bin.
Here, $\kappa_{\rm th}$ denotes the threshold applied to the predicted
concentration parameter.

\begin{figure}
    \centering
    \includegraphics[width=0.9\linewidth]{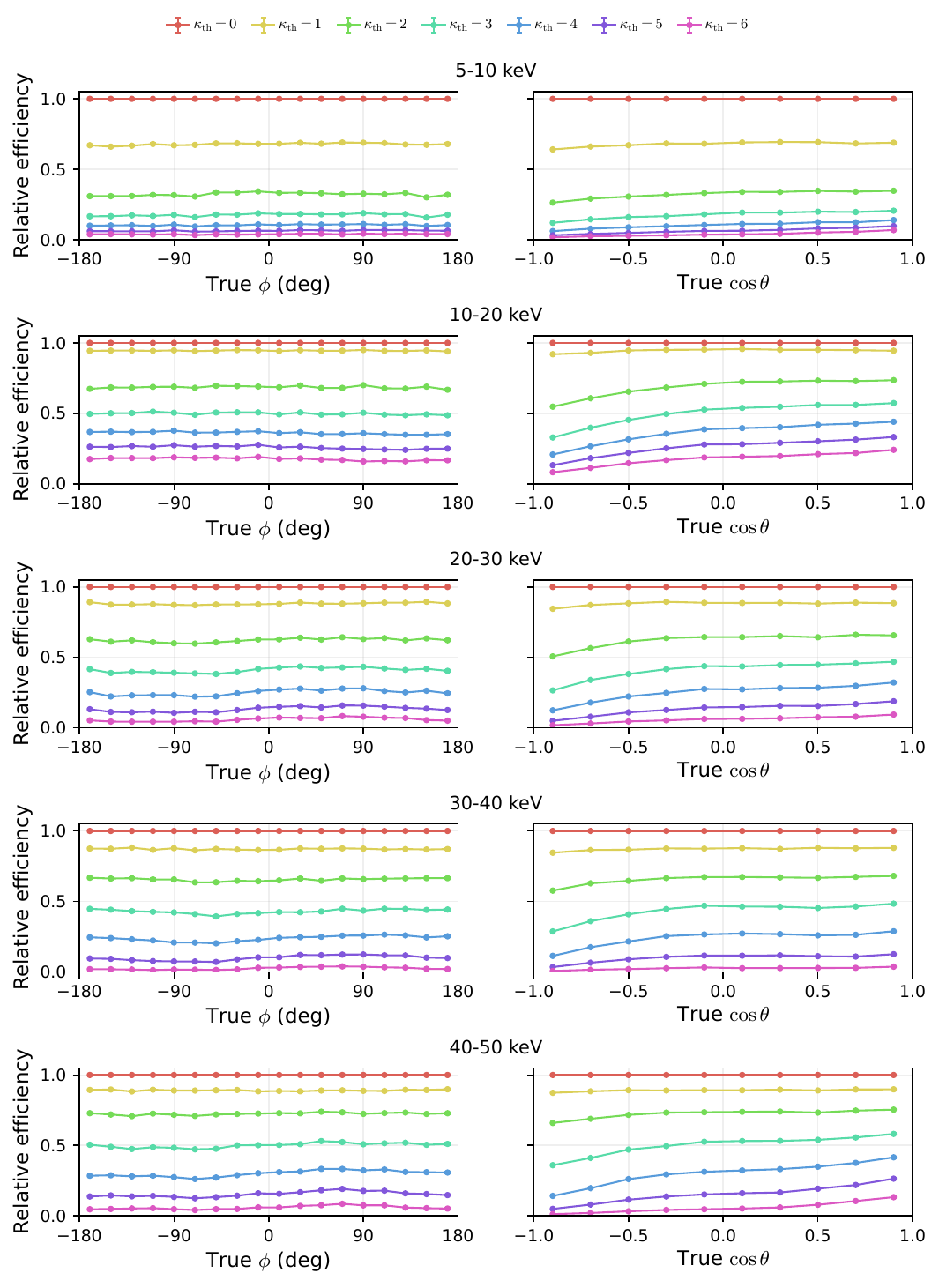}
    \caption{
    Directional dependence of the relative efficiency after applying different
thresholds $\kappa_{\rm th}$.
The left and right panels show the relative efficiency as a function of the
true azimuthal angle $\phi$ and true $\cos\theta$, respectively, for the five
energy ranges from 5--10 to 40--50~keV.
    }
    \label{fig:kappa_direction_efficiency}
\end{figure}

Figure~\ref{fig:kappa_direction_efficiency} shows the relative efficiency as a function of the true $\phi$ and $\cos\theta$ for different values of $\kappa_{\rm th}$.
The efficiency is approximately uniform with respect to $\phi$ over
the investigated energy range.
In contrast, a dependence on $\cos\theta$ becomes apparent as $\kappa_{\rm th}$ is increased, with the relative efficiency decreasing toward $\cos\theta=-1$.
These results indicate that the $\kappa$-based selection does not
introduce a significant azimuthal dependence, whereas a stringent
selection introduces a direction-dependent acceptance along the
detector $z$ axis. 

\end{document}